\documentclass[twocolumn,prl,twocolumn]{revtex4}
\usepackage[latin9]{inputenc}
\usepackage{color}
\usepackage{amsmath}
\usepackage{amssymb}
\usepackage{graphicx}
\usepackage{esint}
\usepackage[unicode=true,
 bookmarks=true,bookmarksnumbered=false,bookmarksopen=false,
 breaklinks=false,pdfborder={0 0 1},backref=false,colorlinks=true]
 {hyperref}
\hypersetup{
 linkcolor=magenta, urlcolor=blue, citecolor=blue, pdfstartview={FitH}, hyperfootnotes=false, unicode=true}

\makeatletter
\@ifundefined{textcolor}{}
{%
 \definecolor{BLACK}{gray}{0}
 \definecolor{WHITE}{gray}{1}
 \definecolor{RED}{rgb}{1,0,0}
 \definecolor{GREEN}{rgb}{0,1,0}
 \definecolor{BLUE}{rgb}{0,0,1}
 \definecolor{CYAN}{cmyk}{1,0,0,0}
 \definecolor{MAGENTA}{cmyk}{0,1,0,0}
 \definecolor{YELLOW}{cmyk}{0,0,1,0}
}

\usepackage{amsfonts}\usepackage{tabularx}\usepackage{dcolumn}\usepackage{bm}

\usepackage[outdir=./]{epstopdf}

\hypersetup{urlcolor=blue}

\makeatother

\begin{document}

\title{Exponential Orthogonality Catastrophe in Single-Particle and Many-Body Localized
Systems}

\author{Dong-Ling Deng, J. H. Pixley, Xiaopeng Li, and S. Das Sarma}

\affiliation{Condensed Matter Theory Center and Joint Quantum Institute, Department
of Physics, University of Maryland, College Park, MD 20742-4111, USA}
\begin{abstract}
We investigate the statistical orthogonality catastrophe (StOC) in single-particle and many-body localized systems by studying the response of the many-body ground state to a
 local quench. Using scaling arguments and exact numerical calculations,
we establish that the StOC gives rise to a wave function overlap between
 the pre- and post-quench ground states
that has an \emph{exponential}
decay with the system size,
 in sharp contrast to the well-known power law Anderson orthogonality catastrophe in metallic systems. This exponential decay arises from a statistical charge transfer process where a particle can be effectively ``transported'' to an arbitrary lattice site.
 In  a many-body localized phase, this non-local transport and the associated exponential StOC phenomenon persist in the presence of interactions.
 We study experimental consequences of the exponential StOC on  Loschmidt echo and spectral function, establishing that this phenomenon should be observable
in cold atomic experiments through 
Ramsey interference and radio-frequency spectroscopy.

\end{abstract}

\maketitle



The response of many-body states to local quenches can be remarkably
drastic; even an arbitrarily weak local perturbation
can substantially
modify the structure of the final ground state. As shown in the seminal
work by Anderson \cite{Anderson1967Infrared}, the overlap between
the ground states of a metallic (i.e. extended states) system with ($|G'\rangle$) and without ($|G\rangle$) a local perturbation
has a power-law decay in the system size $F\equiv|\langle G|G'\rangle|\sim L^{-\gamma}$,
where $\gamma>0$ is a function of the scattering phase shift created
by the local quench, and thus the overlap vanishes in the thermodynamic limit (and hence an infrared `catastrophe'). 
This phenomenon
is the celebrated Anderson orthogonality catastrophe (AOC) \cite{mahan2000many},
which has far-reaching
consequences
in various contexts,
ranging from the x-ray edge singularity \cite{mahan2000many} and Kondo effect
\cite{Hentschel2007Orthogonality,Yuval1970Exact,Tureci2011Many-body,Latta2011quantum,Munder2012Anderson}
to Luttinger liquids \cite{Meden1998Orthogonality,Gogolin1993Local,Pustilnik2006Dynamic,Imambekov2012One,Imambekov2009universal}
and resonant tunneling in mesoscopic structures \cite{Matveev1992Interaction,Abanin2005Fermi,dAmrumenil2005Fermi,Hentschel2005Fermi,Geim1994Fermi,Abanin2004Tunable}. Moreover, a power-law multifractal orthogonality catastrophe (OC) can even occur at random singlet quantum critical points~\cite{Vasseur2015multifractal}. 


Quantum localized systems, both single-particle \cite{Anderson1958Absence}
and many-body localized (MBL) \cite{Basko2006metal,Gornyi2005Interacting,altman2015universal,Nandkishore2015many},
show a cornucopia of intriguing properties, such as violation of
ergodicity and emergent integrability \cite{Rosa2015integrals,Huse2014Phenomenology,Serbyn2013Local}.
MBL has attracted tremendous attention recently due to the breakdown of conventional quantum statistical mechanics and the existence of infinite-temperature
dynamical quantum phase transitions~\cite{Pal2010Manybody} that appear to defy any
description within a conventional formalism~\cite{Potter2015Universal,Vosk2015Theory}.
 Despite extensive research in the subject,
the orthogonality catastrophe, a compelling phenomenon well studied in diffusive metals~\cite{Chen1992Xray,Aleiner1998Shifts,mahan2000many},  remains
largely unexplored for localized systems \cite{Gefen2002Anderson}.
Very recently, in an important paper~\cite{Khemani2014nonlocal}, a new idea of statistical orthogonality catastrophe (StOC) has been introduced for the  one dimensional (1D) localized Anderson model.  

\begin{figure}
\includegraphics[width=0.46\textwidth]{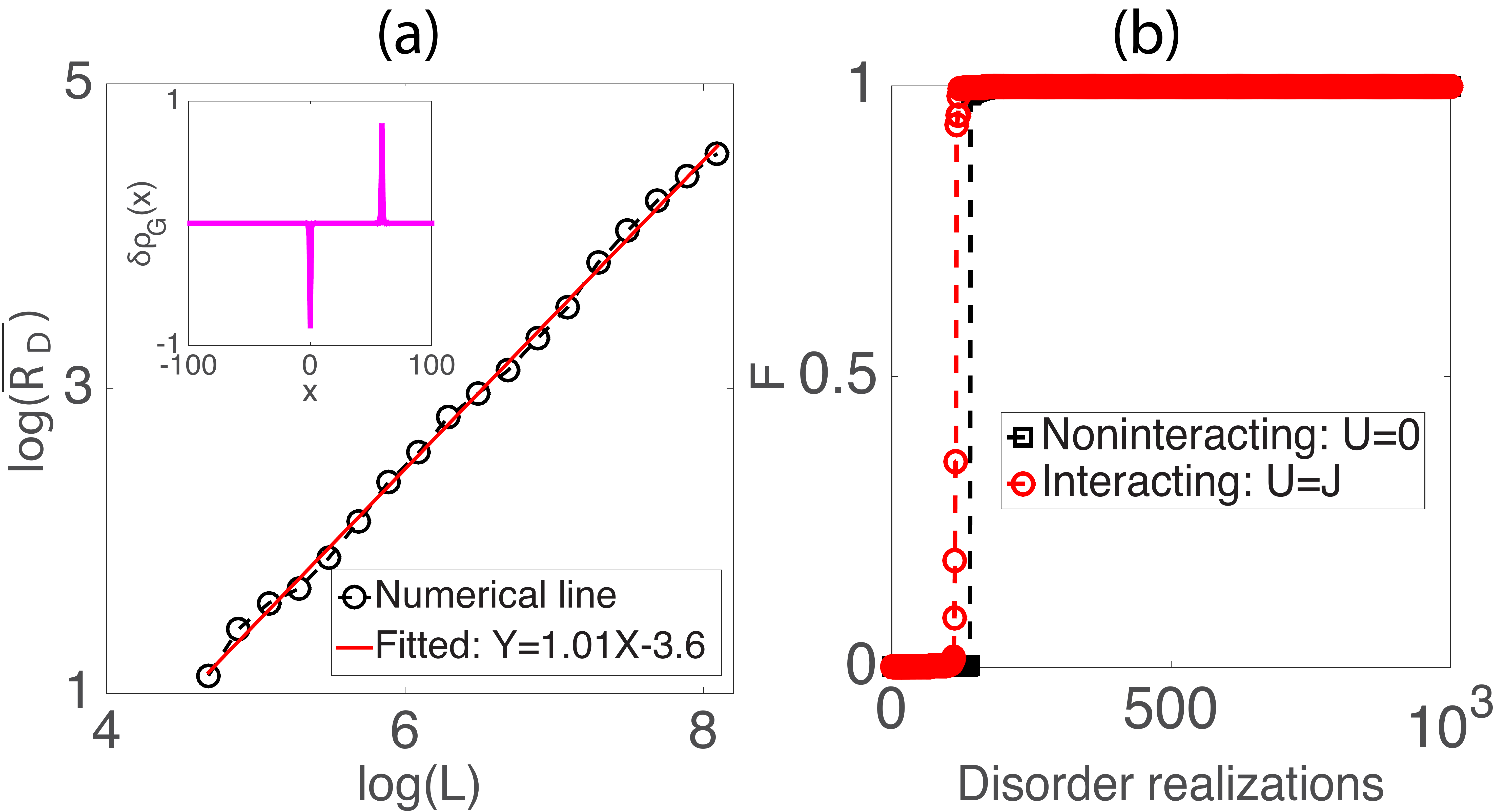}
\protect\caption{ Nonlocal charge transfer and StOC for the AA model.
(a) Scaling of the radius of zone of disturbance
($\overline{R_D}$) with system size in the non-interacting AA model.  A linear scaling $\overline{R_{D}}\sim0.03L$
is obtained. 
To suppress statistical errors, we averaged over $10^{4}$ different values of $\phi$ (evenly distributed in $[-\pi,\pi]$).
Inset: nonlocal adiabatic charge transfer for a fixed $\phi=5\pi/4$.
A particle on site $0$ is transferred to a distant location (near
site $60$) by the time-dependent local potential $V_{0}$.
(b)
Many-body ground state overlaps sorted over $10^{3}$ different values
of $\phi$ for the AA model with and without interaction. Depending on whether there is a charge transfer or not
during the adiabatic process, the overlap $F$ is vanishing or close
to unity, leading to a `statistical' version of OC with probability $P$
around $0.25$.
Here we fix $v_{0}=0.4\lambda=1.6J$.
}
\label{fig:Nonlocal-charge-transfer}
\end{figure}

\begin{figure*}
\includegraphics[width=0.83\textwidth]{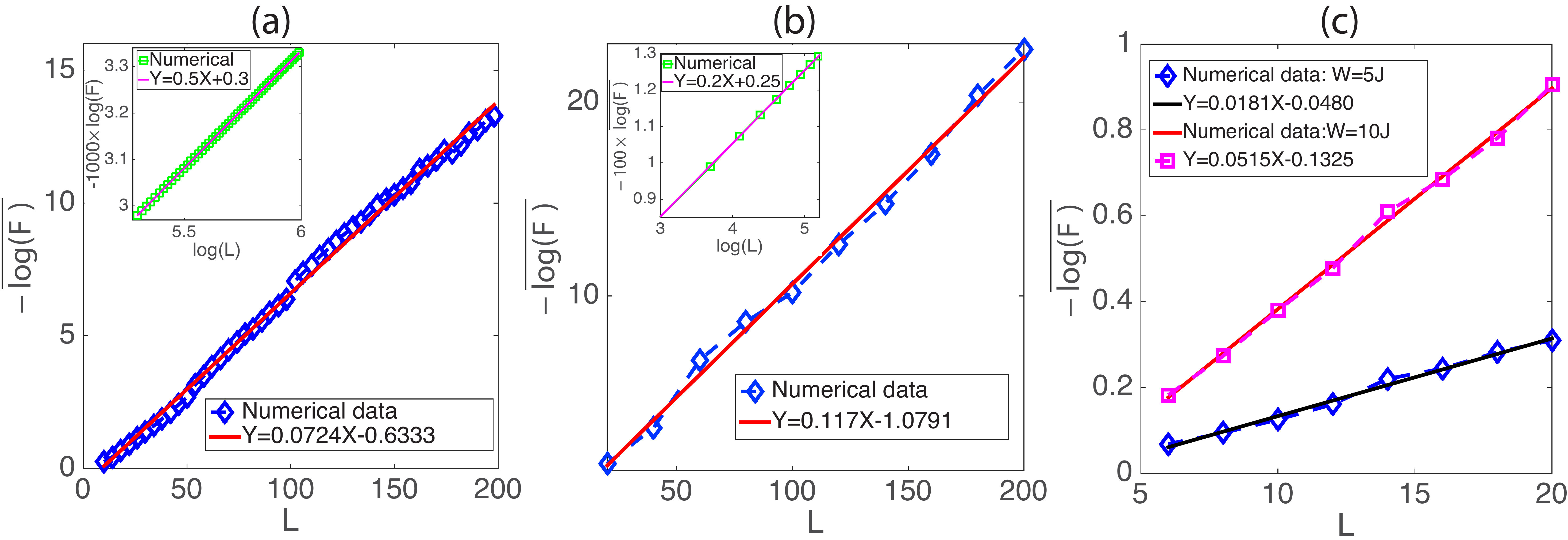}

\protect\caption{Exponential StOC in single-particle and many-body localized
systems. (a) Anderson
model with $v_{0}=0.4W=4J$ and $U=0$. $\overline{\log(F)}$ is
obtained by averaging over $10^{5}$ disorder realizations. The typical
fidelity has an exponential decay $F_{\text{typ}}\sim e^{-0.0724L}$.
Inset: AOC for a clean (metallic) system with $W=0$, $v_{0}=0.2J$
and $U=0$. The overlap has a power-law decay $F\sim L^{-5\times10^{-4}}$,
as expected from Anderson \cite{Anderson1967Infrared}. (b) AA model with $v_{0}=0.4\lambda=1.6J$
and $U=0$. $\overline{\log(F)}$ is obtained by averaging over $10^{5}$
different values of $\phi$, which are uniformly distributed in $[-\pi,\pi].$
The typical fidelity has an exponential decay $F_{\text{typ}}\sim e^{-0.117L}$.
Inset: OC with power-decaying overlap $F_{\text{typ}}\sim L^{2\times10^{-3}}$
in the delocalized region with $\lambda=0.25v_{0}=0.1J$. (c)  
Anderson model a MBL phase with $U=J$ and $v_{0}=0.4W$. $\overline{\log(F)}$
is obtained by averaging over $10^{4}$ disorder realizations. The
typical fidelity has a exponential decay $F_{\text{typ}}\sim e^{-0.0515L}$
($F_{\text{typ}}\sim e^{-0.0181L})$ for $W=10J$ ($W=5J$).
\label{fig:ExpOC}}
\end{figure*}

Here, we study the StOC in single-particle and many-body localized systems  (see Fig.~\ref{fig:Nonlocal-charge-transfer}).
Through scaling arguments and exact diagonalization, we show
that StOC gives rise to a wave function overlap that is \emph{exponentially} suppressed in the system size and is much stronger, albeit statistical in nature,
 than
the well-studied power-law AOC in extended  systems.
%
%
In particular, by studying both  1D Anderson and Aubry-Andre (AA) models (both with and without interactions) we show that the exponential StOC
 is a generic many-body phenomenon in localized systems
(both single-particle and many-body localized) with a typical wave function overlap $F_{\text{typ}}  \equiv \exp{(\overline{\log F})}$  going as
\begin{eqnarray}
F_{\text{typ}}  & \sim \exp{(-\beta L)},
\label{eq:ExponentialLaw}
\end{eqnarray}
as shown in Fig.~\ref{fig:ExpOC}. Here, $\beta>0$ is related to
the localization length and the strength of the local quench. Moreover, by using the kernel polynomial method \cite{WeiSe2006The},  we calculate the spectral function and Loschmidt echo, which allows us to demonstrate that they manifest certain unique features that are related to exponential StOC and can be directly observed in a cold atom experiment through Ramsey interference or radio-frequency spectroscopy \cite{Knap2012Time}.  Since the fundamental nature of Anderson and AA localizations are qualitatively different, with quantum interference and quantum gap spectrum playing the key role respectively, we hypothesize that all localized systems (single-particle or many-body) are characterized by the exponential StOC established in the current work.



\textit{Model system}.---
In this work,
we consider a class of 1D interacting fermionic 
initial Hamiltonians defined as
\begin{eqnarray}
H_{I}=\sum_{j=-L/2+1}^{L/2}-J(a_{j}^{\dagger}a_{j+1}+h.c.)+\mu_{j}n_{j}+Un_{j}n_{j+1}, \,\,\,
\label{eq:Hamiltonian-AI-AA}
\end{eqnarray}
where $a_{j}$ ($a_{j}^{\dagger}$) are fermionic annihilation (creation)
operators, $n_{j}=a_{j}^{\dagger}a_{j}$ is the corresponding fermion
number operator, $J$ is the nearest-neighbour hopping strength, and
$U>0$ is the nearest-neighbor interaction.
Depending on how
we choose the local chemical potential, the model in Eq. (\ref{eq:Hamiltonian-AI-AA})
can be reduced to either the  Anderson model~\cite{Anderson1958Absence}
if the local potential $\mu_{i}$ is drawn from a uniform random distribution between $[-W,W]$, or the
AA model~\cite{aubry1980analyticity,Iyer2013Many-body}
if 
$\mu_{j}=\lambda\cos(2\pi\alpha j+\phi)$, 
where
$\lambda$ is the strength of the potential, with $\alpha$ an irrational number and $\phi$ a random phase. Both Anderson and AA models have been extensively
studied in the context of many-body localization (see Ref. \cite{Abrahams1979Scaling,Pal2010Manybody,bera2015many}). To unify the notation between the two models we will use $g$ to denote $W$ and $\lambda$ respectively for the 1D Anderson and AA model.
In both cases, we start
 with a localized many-body ground state $|G\rangle$,
we then quench the model  by introducing 
a local impurity
into
 the
Hamiltonian through an onsite potential $V_{0}=v_{0}a_{0}^{\dagger}a_{0}$ ($v_0>0$) at site $0$ and 
arrive at
 a final Hamiltonian
$H_{F}=H_{I}+V_{0}$, with a 
new ground state $|G'\rangle$.
We investigate the asymptotic behavior
of the many-body ground
state
overlap (fidelity) defined by $F\equiv|\langle G|G'\rangle|$,
as a function of the system size.

\emph{Charge transfer}---
We consider a time-dependent Hamiltonian $H(t)=H_{I}+V_{0}(t/\tau)$,
where $H_{I}$ is a static Hamiltonian describing a localized system
and $V_{0}(t/\tau)$ a time-dependent local potential at site $0$, and $\tau$ is chosen large enough so that adiabaticity is maintained. The time dependence of the local potential is $V_{0}(-\infty)=0$ and $V_{0}(\infty)=v_{0}a_{0}^{\dagger}a_{0}$.
In order to characterize the
 charge response,
we define the
change of density as $\delta\rho(x)  \equiv  \langle n_{x}\rangle_{t=\infty}-\langle n_{x}\rangle_{t=-\infty}$, 
where $\langle n_{x}\rangle_{t}=\langle G(t)|n_{x}|G(t)\rangle$,
$|G(t=-\infty)\rangle=|G\rangle$ and $|G(t=\infty)\rangle=|G'\rangle$ denote respectively the instantaneous initial and final ground states of $H(t)$~\cite{Khemani2014nonlocal}.
As a result of the local potential, for certain disorder realizations such that $\mu _0<0$, $\mu_0+v_0>0$
charge will be transferred from site $0$ to an arbitrary site $R_F$ (see the inset of Fig.~\ref{fig:Nonlocal-charge-transfer} (a)),
 with $1<|R_F|\le L/2$ (see our coordinate choice in Eq.~\eqref{eq:Hamiltonian-AI-AA}),
and therefore the initial and final ground state wave functions will be orthogonal ($F\rightarrow 0$
 in the thermodynamic limit).
Whereas for other realizations
no charge transfer will take place and therefore $F\approx 1$. It is in this sense that the StOC in localized systems is \textit{statistical}, depending on specific random potential realizations. This is captured in our numerical calculations for the AA model with and without interactions as shown
 in Fig. \ref{fig:Nonlocal-charge-transfer}(b),
where we plot $F$
sorted in magnitude over $10^{3}$
different random realizations of $\phi$. For $U=0$, we find an OC with probability
$P(-v_0<\mu_0<0)\approx0.25$, whereas for $U/J=1$ the probability is reduced slightly.
We also establish  StOC for the interacting Anderson model \cite{supplement}.

 In order to quantify the distance the charge will move,
we consider the so-called radius
of the ``zone of disturbance'' over which charge transfer takes
place
\begin{eqnarray}
R_{D} & \equiv & \frac{\int_{-L/2}^{L/2}|x\delta\rho(x)|dx}{\int_{-L/2}^{L/2}|\delta\rho(x)|dx}.
\label{eq:R1DistZone}
\end{eqnarray}
As a result of the charge transfer landing on any arbitrary site the disorder average of $R_D$
can
have a very wide spatial distribution, giving rise to the scaling $\overline{R_{D}}\sim\eta(g,U) L$,
as shown in Fig.~\ref{fig:Nonlocal-charge-transfer} (a) for the non-interacting AA model, in agreement with Ref.~\cite{Khemani2014nonlocal} for the 1D non-interacting Anderson model.
In addition, we also establish $\overline{R_{D}}\sim L$ in the MBL phase of both interacting 1D Anderson and AA models \cite{supplement}. 
Thus, we have established $\overline{R_D}\sim L$
 for the ground state of generic localized systems, irrespective of single-particle or many-body localized.

\textit{Finite size scaling of $F$}---
Here, we establish the leading finite size scaling of $F$ in the non-interacting limit
 based on the dynamic charge transfer
picture presented above. We then generalize the scenario to the MBL case.
We begin by considering $N$ particles in the localized phase. The many-body ground state is constructed by filling the $N$ lowest single-particle eigenstates, which have localized wave functions going as $\phi_n(x) \propto \exp(-|x-x_n|/\xi)$, with $x_n$ the localization center and $\xi$  the single-particle localization length. Focusing on a disorder realization that transfers charge from site $0$ to site $R_f$, the leading contribution to the overlap goes as
\begin{eqnarray}
F & \sim & \int dx \phi_0(x)\phi_f '(x)  \sim e^{-|R_{f}|/\xi},
\end{eqnarray}
 with $\phi_f ' (x)$ referring to the post-quench single-particle wave function.
Consequently, 
the disorder averaged $\overline{\log F}\sim-\overline{|R_{f}|}/\xi$, where
$\overline{|R_{f}|}$ is the averaged distance
of charge transfer,
which has
the same scaling
as $\overline{R_{D}}$,  
i.e., $\overline{|R_{f}|}\sim\overline{R_{D}}\sim L$. Thus, we arrive at 
one of our main results
$\overline{\log F}\sim-\beta L$,
and
\begin{equation}
F_{\text{typ}}\sim \exp{(-\beta L)}, \,\,\,\,\,\, \beta = P/4\xi,
\end{equation}
where $P$ is the probability of creating the charge transfer event, i.e. $P=v_0/(2W)$ and $P(-v_0<\mu_0<0)$ for the non-interacting Anderson and AA models, respectively.

The above argument can also carry over to the MBL case
 via the `l-bits' formalism where the `l-bits' are related to charge degrees of freedom by quasi-local unitary transformations
and their interactions are exponentially local~\cite{Serbyn2013Local,Huse2014Phenomenology}.
We then expect  from the charge transfer picture
that the overlap between interacting many-body ground states before and after the charge transfer to go as $F\sim \exp(-|R_f|/\xi_{\mathrm{MBL}})$, where $\xi_{\mathrm{MBL}}$ is the many-body localization length that can be computed through local integrals of motion~\cite{Chandran2015Constructing}. Consequently  for this many-body case, $F_{\mathrm{typ}}$ still approaches zero exponentially as the non-interacting case, yet with $\beta = P_{U\neq 0}/\xi_{\mathrm{MBL}}$.
We now turn to exact numerical calculations to put the scaling analysis on a solid footing.

\textit{Numerical results.}---
We now  present our
 numerical results in order to establish the exponential StOC for localized systems.
We use exact diagonalization to diagonalize
$H_{I}$ and $H_{F}$ to obtain the energy spectrum and eigenstates.
For the noninteracting case, the many-body ground states are constructed
by simply filling the single-particle eigenstates and the overlaps
are calculated via determinants (see Supplemental Material \cite{supplement}).
For the interacting case, the overlaps can be computed directly by
inner products of $|G\rangle$ and $|G'\rangle$ after an exact diagonalization of the many-body hamiltonian. We then calculate the
wave function overlap ($F$) averaged over 
different disorder realizations
for both
Anderson and AA models.

\begin{figure}
\includegraphics[width=0.484\textwidth]{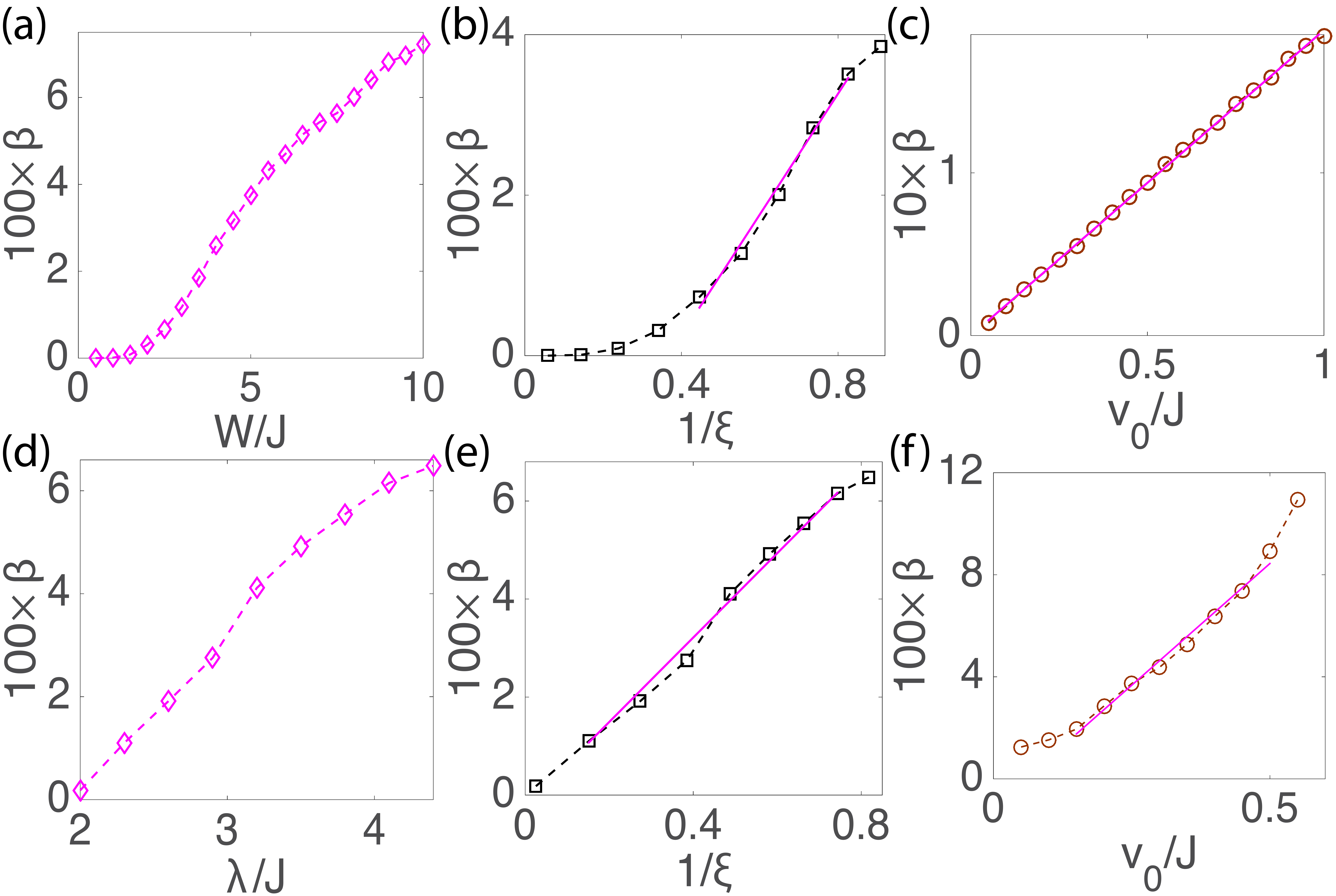}

\protect\caption{$\beta$ dependences for the  Anderson ((a), (b), and (c))
and AA ((d), (e), and (f)) models in the noninteracting limit. The other parameters are chosen as $v_0=0.4W$ ($v_0=0.4\lambda$) in (a) and (b) ((d) and (e)); $W=10J$ in (c) and $\lambda=4J$ in (f). Dashed magenta lines are fits to the linear relationships in each case.
\label{fig:BetaTrend}}
\end{figure}

 Our numerical results for the StOC in the Anderson and AA models are shown in Fig.~\ref{fig:ExpOC}, 
which clearly confirms our scaling analysis, and establishes the exponential StOC $F_{\mathrm{typ}}\sim \exp(-\beta L)$.
For the sake of comparison,
we have also plotted the results for the clean (metallic, $W=0$) limit of the
Anderson model and for the delocalized phase of the AA model ($\lambda/J <2$),
respectively. For these extended states, the overlaps follow a power-law decay,
as expected from AOC~\cite{Anderson1967Infrared}.  For the non-interacting case we are able to reach sufficiently large system sizes, whereas in the MBL phase we are limited to $L=24$ as our largest system.

In Fig. \ref{fig:BetaTrend}, we
 show
the dependence of $\beta$ on $W,\lambda,\xi$ and $v_0$ in the non-interacting limit.
First, we focus on the strength of the on site potential,
 which is related to the localization length as $1/\xi \sim | g- g_c|^{\nu}$ near $g_c$ with a localization length exponent $\nu$
 (for the 1D Anderson and AA models, $g_c$ is $0$ and  $2\lambda$,  respectively).
Deep in the localized phase $\xi$ saturates to the lattice spacing,
 whereas
it diverges on  approaching the localization transition point. We thus expect
$\beta$ to be zero at $g = g_c$ and to increase as we increase $g$ from $g_c$,
which is in excellent agreement with our numerical calculations. In addition, the scaling analysis also predicts $\beta \propto 1/\xi$. We compute the localization length directly from the decay of the single-particle wavefunction $\phi(x)$ from its maximal value (in absolute value) and then average over disorder realizations.  Close to the localization transition and deep within the localized phase such a method of calculating $\xi$ suffers from large numerical error, nonetheless in the intermediate parameter regime we establish a clear linear
 relation.
Finally, we come to the dependence of $\beta$ on $v_0$ for a fixed $g$.
 The
results are in good agreement with $\beta \propto v_0$, which allows us to conclude that our numerical results are consistent with $\beta = P/4\xi$, and thus provide another independent confirmation of our scaling analysis.

\begin{figure}
\includegraphics[width=0.48\textwidth]{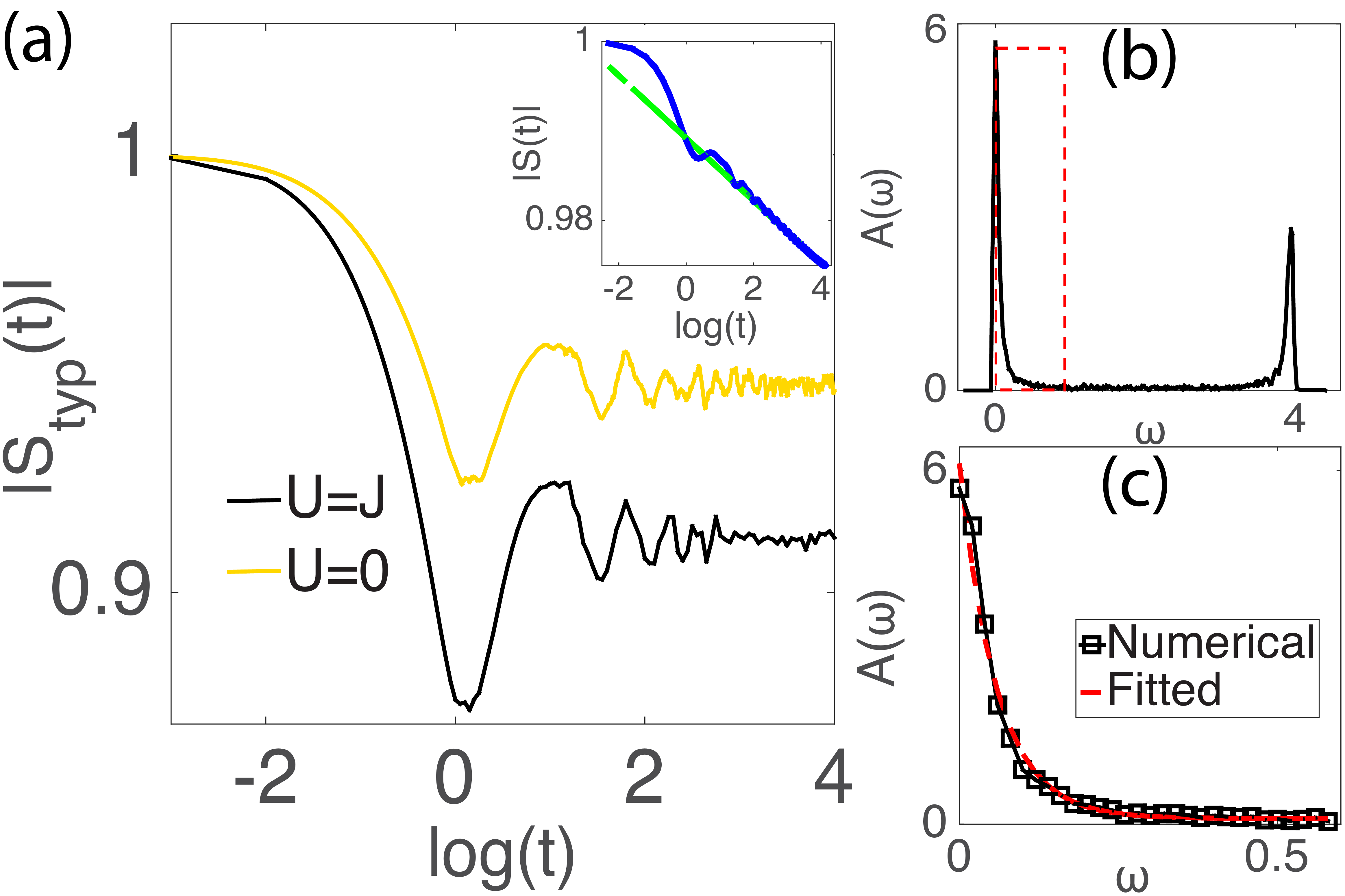}

\protect\caption{Numerical simulations of the typical Loschmidt echo and spectral functions. (a) The typical Loschmidt echo as a function of time for the Anderson model with and without interaction. Inset: Loschmidt echo in the  clean limit, which has a power-law decay at long time. (b) The disorder-averaged spectral function for the interacting Anderson model. Unlike the metallic or the ordinary band-insulator cases, $A(\omega)$ for localized systems has two divergent peaks at either $\omega=0$ or $\omega=v_0$ and non-zero weight between $(0,v_0)$. (c) Zooming in on the red rectangular region of (b). Near $\omega=0$, an exponential decay $A(\omega)\sim 6\times e^{-17.1\omega}+0.1$ is obtained. The parameters are chosen the same as in Fig.\ref{fig:ExpOC} (see \cite{supplement} for more details). 
\label{fig:LosmidEAw}}
\end{figure}

\textit{Experimental detection}. ---The AOC has far-reaching consequences for several dynamical phenomena, such as X-ray edge singularity in metals~\cite{mahan2000many} and power-law suppressed Loschmidt echo for ultracold atoms in the long time limit \cite{Knap2012Time-dependent}. 
 We expect the exponential StOC in localized systems to  have significant implications in similar experimental settings.  
 In a  setup proposed to probe the OC physics \cite{Knap2012Time-dependent,Goold2011Orthogonality,Sindona2013Orthogonality}, 
a nature experimentally measurable quantity to characterize the exponential StOC is the typical Loschmidt echo defined as $S_\text{typ}(t)\equiv \text{exp} (\overline{\log |S(t)|})$, where   
\begin{eqnarray}
S(t)=\langle G| e^{iH_It}e^{-iH_Ft}| G\rangle.
\end{eqnarray} 
In Fig. \ref{fig:LosmidEAw} (a) we show $S_\text{typ}(t)$ for the Anderson model \cite{Footnote}.  As a result of the StOC and the localized nature of the many body ground state we find $S_\text{typ}(t)$ 
 saturates to a  
 nonzero value in the long-time limit.
This is in sharp contrast to the
metallic case where $|S(t)|$ has a power-law decay due to the AOC \cite{Knap2012Time-dependent}. 
Another quantity that is also directly measurable in experiments is the spectral function (which is related to the ``hole" Green function in the AOC literature \cite{mahan2000many})  $A(\omega)=\sum_{n'} |\langle G|n'\rangle|^2\delta(\omega-(E_{n'}-E_G))$, where $|n'\rangle$ is the $n'$-th eigenstate of $H_F$ with eigen energy $E_{n'}$ and $E_G$ is the ground state energy of $H_I$. Using the kernel polynomial method \cite{WeiSe2006The}, we calculate the disorder-averaged spectral function for both the AA and Anderson models (see \cite{supplement}). 
 As shown in Fig. \ref{fig:LosmidEAw}(b) and (c),  the spectral function exhibits two 
 peaks with exponentially decaying tails, which are unique features of the exponential StOC in localized systems (whereas in the AOC, $A(\omega)$ has only one peak and a power-law dependence \cite{mahan2000many}).

With the recent experimental realizations of single-particle \cite{Billy2008direct,Kondov2011three,Roati2008anderson}  and many-body localization \cite{Kondov2015Disorder,Schreiber2015observation,Smith2015Many}, the above intriguing features could be experimentally observed with 
 present techniques. 
Considering a ultracold atom experiment for instance, the localized ground states could be achieved by 
 adiabatic ramping 
and the local quench can be switched on and off by manipulating locally two-level atoms (the impurity atoms of a strongly imbalanced mixtures \cite{Zwierlein2006Fermionic,Liao2010Spin}), and the Loschmidt echo and the spectral function can be directly measured through Ramsey interference and radio-frequency spectroscopy, respectively (see Ref. \cite{Knap2012Time-dependent,Goold2011Orthogonality,Sindona2013Orthogonality} for details). 
 In electronic materials, 
nontrivial effects of the exponential StOC may have already been observed  
in recent experiments of quantum electron glasses \cite{Ovadyahu2015Infrared,Ovadyahu2007Relaxation}, where dramatic sluggish dynamics have been reported.

To conclude, we have established a statistical exponential orthogonality catastrophe in both single-particle and many-body localized systems. We introduced scaling arguments to explain this `stronger' catastrophe.  Using  exact numerical calculations we have verified the scaling analysis and its generalization to the MBL case. The exponential StOC can be used as a diagnostic tool to identify localization and could be directly observed in cold-atomic or other mesoscopic systems with current technologies.

\begin{acknowledgments}
We thank S.-T. Wang, Z.-X. Gong, Y.-L. Wu, and J. D. Sau for discussions. This
work is supported by LPS-CMTC, JQI-NSF-PFC, and ARO-Atomtronics-MURI. The authors acknowledge the University of Maryland supercomputing resources (http://www.it.umd.edu/hpcc) made available in conducting the research reported in this paper.
\end{acknowledgments}

\bibliographystyle{apsrev4-1}


%

\begin{widetext}

\section{Supplemental Material for ``Exponential Orthogonality Catastrophe in Single-particle and Many-body Localized
Systems''}

\subsection{Calculations of ground-state overlaps for free fermions}

In this section, we provide the details on how we calculate the ground-state
overlap for free fermions. As in the main text, we denote by $|G\rangle$and
$|G'\rangle$ the ground states of $H_{I}$ and $H_{F}$. Since we
only consider the free fermion case, $H_{I}$ ($H_{F}$) can be written
as $H_{I}=\sum_{ij}a_{i}^{\dagger}(\mathcal{H}_{I})_{ij}a_{j}$ ($H_{F}=\sum_{ij}a_{i}^{\dagger}(\mathcal{H}_{F})_{ij}a_{j}$),
where $\mathcal{H}_{I}$ ($\mathcal{H}_{F}$) is the so called kernel
Hamiltonian. One can diagonalize $\mathcal{H}_{I}$ ($\mathcal{H}_{F}$)
by a unitary transformation $\mathcal{U}$ ($\mathcal{V}$): $\mathcal{H}_{I}=\mathcal{U}^{\dagger}\mathcal{E}_{I}\mathcal{U}$
($\mathcal{H}_{F}=\mathcal{V}^{\dagger}\mathcal{E}_{F}\mathcal{V}$)
to find the single-particle eigenmodes $b_{j}=\sum_{k}\mathcal{U}_{jk}a_{k}$
($c_{j}=\sum_{k}\mathcal{V}_{jk}a_{k}$). Here $\mathcal{E}_{I}=\text{diag}(\epsilon_{1}^{I},\epsilon_{2}^{I},\cdots)$
($\mathcal{E}_{F}=\text{diag}(\epsilon_{1}^{F},\epsilon_{2}^{F},\cdots)$)
is a diagonal matrix. Suppose the lattice size is $L$ and the total
particle number is $N$. For free fermionic systems, the ground states
of $H_{I}$ and $H_{F}$ are constructed by filling the lowest $N$
single-particle eigenstates:
\begin{eqnarray*}
|G\rangle & = & \prod_{j}^{N}b_{j}^{\dagger}|0\rangle,\\
|G'\rangle & = & \prod_{j}^{N}c_{j}^{\dagger}|0\rangle.
\end{eqnarray*}
Thus the overlap reads:
\begin{eqnarray}
F & \equiv & |\langle G|G'\rangle|=|\langle0|\prod_{j}^{N}b_{j}\prod_{k}^{N}c_{k}^{\dagger}|0\rangle|.\label{eq:OverlapF}
\end{eqnarray}
Computing $F$ directly from Eq. (\ref{eq:OverlapF}) is challenging.
However, we can simplify Eq. (\ref{eq:OverlapF}) by expressing the
eigenmodes $c_{k}^{\dagger}$s as a combination of $b_{l}^{\dagger}$s
$c_{k}^{\dagger}=\sum_{m}(\mathcal{V}^{\dagger})_{km}a_{m}^{\dagger}=\sum_{m}(\mathcal{V}^{\dagger})_{km}(\sum_{n}\mathcal{U}_{mn}b_{n}^{\dagger})=\sum_{mn}(\mathcal{V}^{\dagger})_{km}\mathcal{U}_{mn}b_{n}^{\dagger}$,
which corresponds to a change of basis. Defining another matrix $A=\mathcal{V}^{\dagger}\mathcal{U}$,
we have $c_{k}^{\dagger}=\sum_{n}A_{kn}b_{n}^{\dagger}$ and
consequently 
\begin{eqnarray*}
F & = & |\langle0|\prod_{j}^{N}b_{j}\prod_{k}^{N}\sum_{n}A_{kn}b_{n}^{\dagger}|0\rangle|.
\end{eqnarray*}
Noting that $\{b_{j},b_{n}^{\dagger}\}=\delta_{jn}$, $b_{n}|0\rangle=0$
and $(b_{n}^{\dagger})^{p}=0$ for all $n$ and $p>1$, we arrive
at a greatly simplified formula 
\begin{eqnarray}
F & = & |\det B|,\label{eq:Fdet}
\end{eqnarray}
where $B_{ij}=A_{ij}$ ($1\leq i,j\leq N$) is a matrix with elements
taken from $A$. The above equation is very useful in our numerical
calculations. With this formula, we are able to calculate the overlap
efficiently for different random realizations.

\subsection{More results on charge transfer and orthogonality catastrophe}

In the main text, we have established an exponential statistical orthogonality
catastrophe in localized systems and its relation to adiabatic charge
transfer. Here we show additional numerical results to support our claim.

\subsubsection{The Anderson model}

In Fig. \ref{fig:OverlapsAndM}, we plot the scaling of $\overline{R_D}$ with system size and the StOC for the 
Anderson model. From Fig. \ref{fig:OverlapsAndM}(a), one indeed obtain a linear scaling $\overline{R_D}\sim L$ for the interacting Anderson model in the MBL phase, thus validating the heuristic arguments in Ref. \cite{Khemani2014nonlocal}. From Fig. \ref{fig:OverlapsAndM}(b), we have statistical orthogonality catastrophe due to the nonlocal charge transfer, as expected.  The probability
of StOC for the noninteracting Anderson model is about $P\approx v_{0}/(2W)=0.2$,
whereas this probability is reduced for the interacting
case, similar to case of the AA model (see Fig. 1 in the main text).

\begin{figure}
\includegraphics[width=0.65\textwidth]{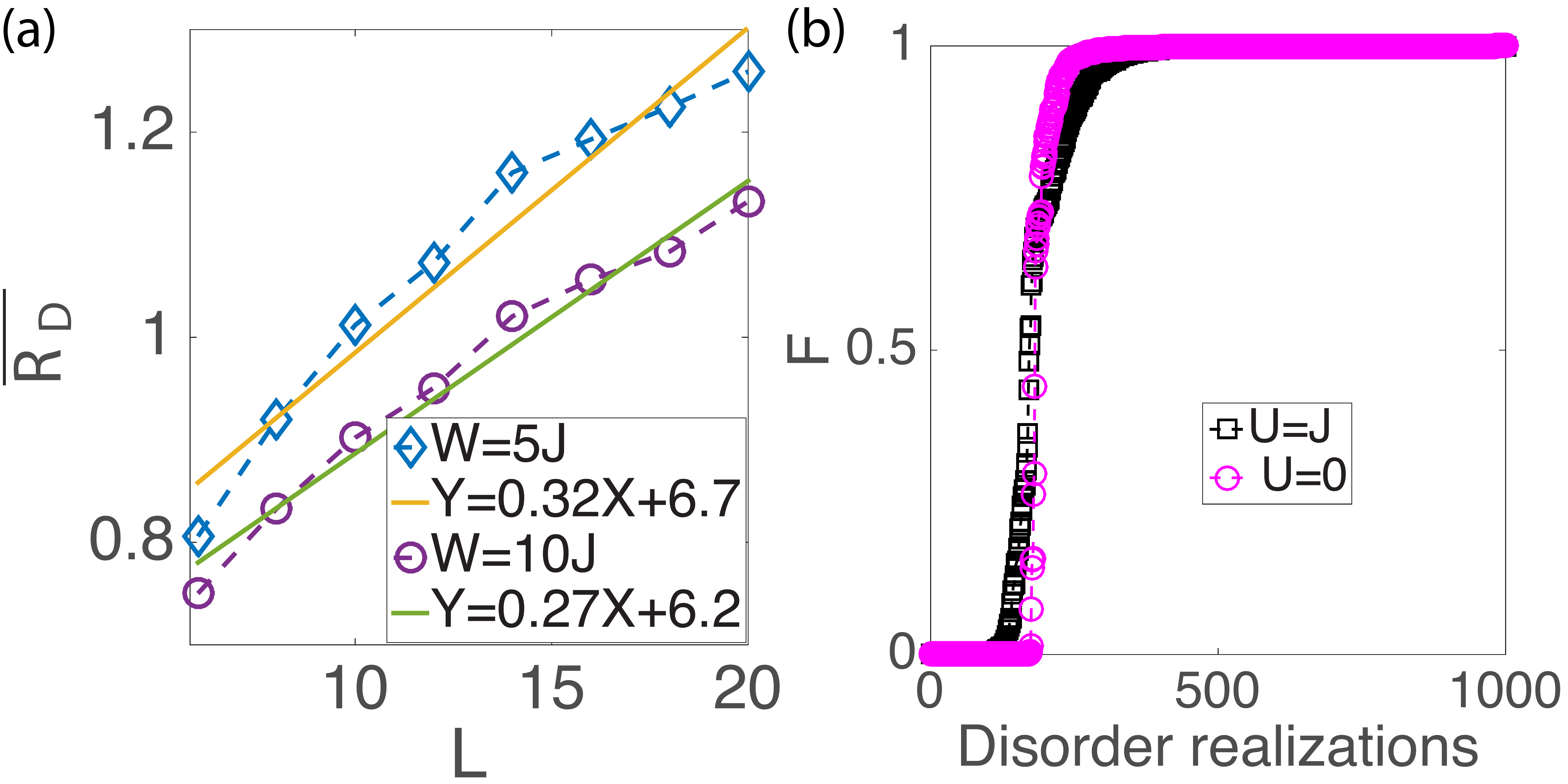}

\protect\caption{Scaling of $\overline{R_D}$ and many-body ground state overlaps for the Anderson model. (a) Scaling of $\overline{R_D}$ with system size for the interacting Anderson model. Linear scalings $\overline{R_D}\sim 0.32L$ and $\overline{R_D}\sim 0.27L$ are obtained for $W=5J$ and $W=10J$, respectively. Here we have fixed $U=J$ and used $10^4$ random realizations to suppress statistical errors. (b) Many-body ground state overlaps sorted over $10^{3}$ disorder realizations for the Anderson model. For the interacting (noninteracting) case, the lattice size is $L=24$
($L=200$). Other parameters are chosen as $v_{0}=0.4W=4J$. \label{fig:OverlapsAndM}}

\end{figure}

\begin{figure}
\includegraphics[width=0.45\textwidth]{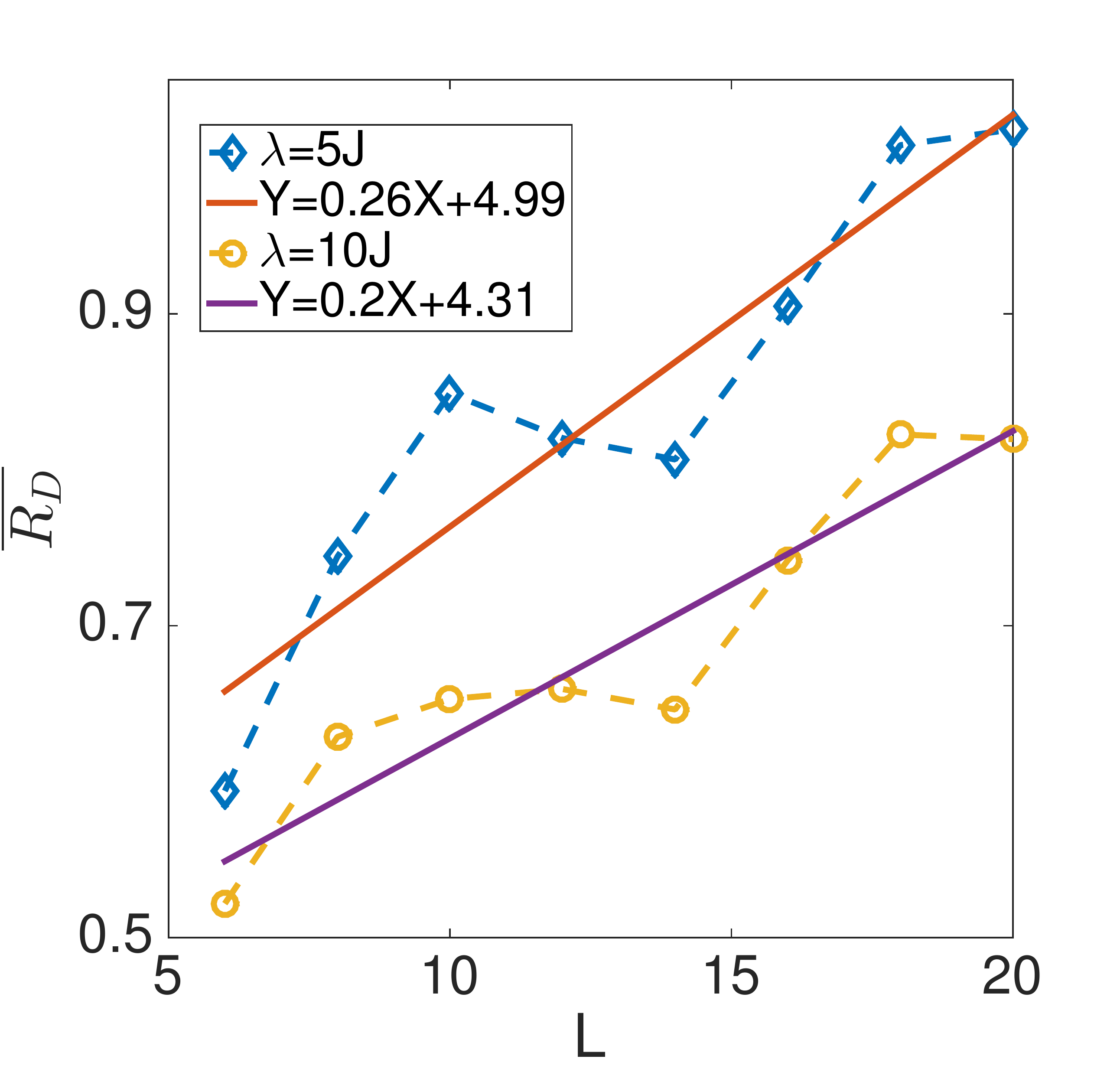}

\protect\caption{Nonlocal charge transfer in the interacting Aubry-Andre model with
$U=J$. $\overline{R_{D}}$ is obtained by averaging over $10^{4}$
different values of $\phi$ evenly distributed in $[-\pi,\pi]$ and
$v_{0}$ is chosen as $v_{0}=0.4\lambda$.\label{fig:ChargeTransAAM}}
\end{figure}

\subsubsection{The Aubry-Andre model}

In the main text, we have shown the StOC in this model for both interacting
and noninteracting cases (see Fig. 1(c)). Here we show more numerical
results on the charge transfer and StOC for the interacting Aubry-Andre
model. In Fig. \ref{fig:ChargeTransAAM}, we plot the scaling of the
radius of zone of disturbance $\overline{R_{D}}$ ($R_{D}$ is defined
by Eq. (4) in the main text) with the system size. 
One may notice
that the numerical lines are not exactly straight lines and there
are large deviations due to the finite-size effect. 
Yet, it is clear
that the qualitative trend is still correct. In Fig. \ref{fig:ExpOCAAM},
we plot the scaling of $-\overline{\log(F)}$ with the system size
for the interacting AA model. Here the finite-size effect is also
non-negligible. In order to clarify the role of finite size effects we compare the many body interacting case with the non-interacting results that can reach much larger systems sizes as shown in Fig. \ref{fig:ComparisonAAM}(a).
%
%
From this figure, it
is clear that the influence of interaction on $-\overline{\log(F)}$
is weak deep in the localized region. Despite the step like features at small $L$, as shown in Fig. \ref{fig:ComparisonAAM}(b)
and (c)  
this is truly just an artifact of small system  sizes, where increasing $L$ gives rise to a clear $-\overline{\log(F)}\sim L$ system size dependence. Thus, as shown in these two figures, the linearity
of $-\overline{\log(F)}$ with $L$ becomes more and more evident
as the system size becomes larger and larger. As we have shown, deep in the localized phase, the presence of interactions
does not quantitatively change $-\overline{\log(F)}$ by a large amount, we thus conclude
that $-\overline{\log(F)}$ also goes linearly with $L$, giving the
exponential scaling $F_{\text{typ}}\sim e^{-\beta L}$ as expected.

\begin{figure}
\includegraphics[width=0.45\textwidth]{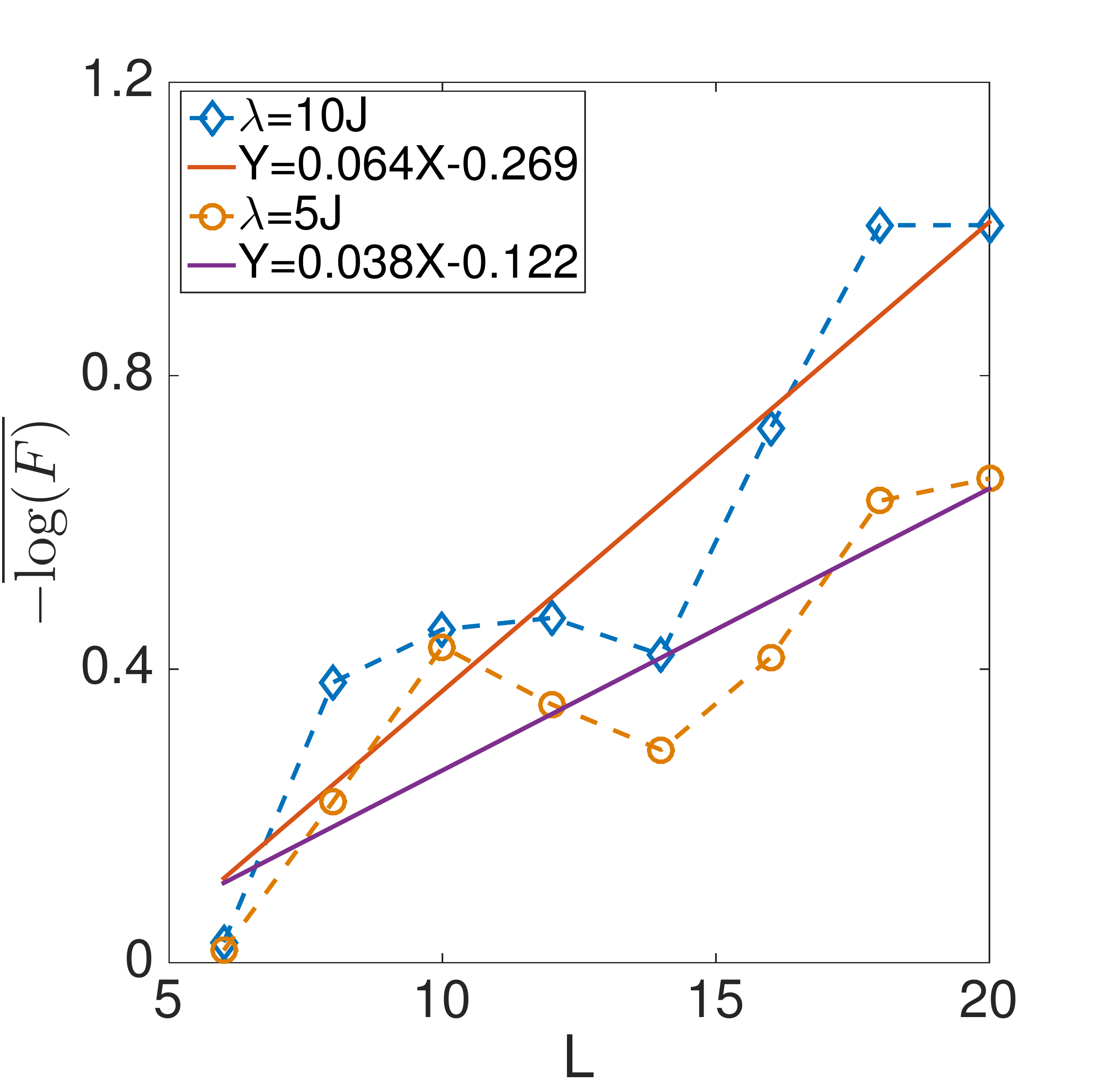}

\protect\caption{Exponential statistical OC in the interacting Aubry-Andre model. Due
to finite size effect, the numerical data has a large deviation from
the corresponding fitted line. $-\overline{\log(F)}$ is obtained
by averaging over $10^{4}$ different values of $\phi$ evenly distributed
in $[-\pi,\pi]$. Other parameters are chosen as $U=J$ and $v_{0}=0.4\lambda$.\label{fig:ExpOCAAM}}
\end{figure}

\begin{figure*}
\includegraphics[width=0.95\textwidth]{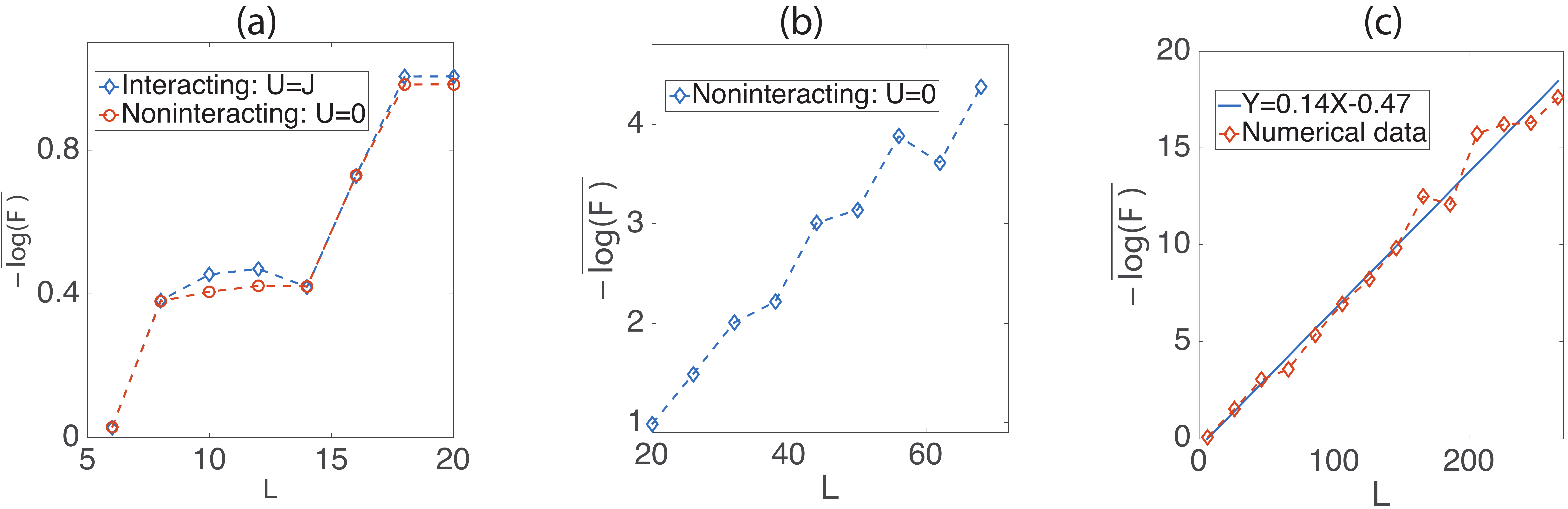}

\protect\caption{(a) Comparison of exponential statistical OC in the interacting and
noninteracting Aubry-Andre model. In the deep localized region ($\lambda=10J$),
the interaction does not modify $-\overline{\log(F)}$ very much.
(b) and (c) plot the scaling of $-\overline{\log(F)}$ with system
size for the noninteracting case. Results are obtained by averaging
over $10^{4}$ different values of $\phi$ evenly distributed in $[-\pi,\pi]$
and here $v_{0}=0.4\lambda=4J$.\label{fig:ComparisonAAM}}

\end{figure*}

\subsection{Calculations of the Loschmidt echo and spectral function}

\subsubsection{Calculations of the Loschmidt echo}
In the main text, we have calculated the Loschmidt echo defined as:
\begin{eqnarray}
S(t)=\langle G|e^{iH_It}e^{-iH_Ft}|G\rangle.\label{LSecho}
\end{eqnarray}
A direct calculation of $S(t)$ using Eq. \ref{LSecho} confronts lots of computational complexity. For the noninteracting cases, the calculation can be greatly simplified. Note that $|G\rangle =\prod_{j}^{N}b_{j}^{\dagger}|0\rangle$ is the ground state of $H_I$, then we have $S(t)=e^{iE_Gt}\langle G|e^{-iH_Ft}|G\rangle$. To calculate $\langle G|e^{-iH_Ft}|G\rangle$, one can do a time-evolution of each eigenmode of $\mathcal{H_I}$
\begin{eqnarray}
b_j(t)=e^{-iH_Ft}b_je^{iH_Ft}=\sum_m\sum_k\left(e^{-i\mathcal{H}_Ft}\right)_{mk}\mathcal{U}_{jk}a_k.
\end{eqnarray}
After  $b_j(t)$ is obtained, we  one can re-express $b_j(t)$ as a combination of $b_l$'s and then the calculation of $S(t)$ is reduced to the computation of a determinant at each time $t$, following the same procedures in Sec. I.  In  Fig. 4(a) of the main text, we have shown the typical Loschmidt echo $S_\text{typ}$ for the noninteracting Anderson model (the yellow curve). In this case, our lattice size is $L=200$ and we used $10^4$ random realizations.  For the interacting case, calculations of $S(t)$ requires full diagonalizations of many-body Hamiltonians and our system size are limited to $L=14$. The numerical results of $S(t)$ for the interacting Anderson model are also shown in Fig. 4(a) (the black curve) in the main text. For both interacting and noninteracting cases, $S(t)$ has various distinctive features that are related to the exponential StOC and are different from that of the Anderson OC for a metallic phase. We have also calculated the Loschmidt echo of the AA model for both interacting and noninteracting cases, and very similar features are obtained.

\begin{figure*}
\includegraphics[width=0.85\textwidth]{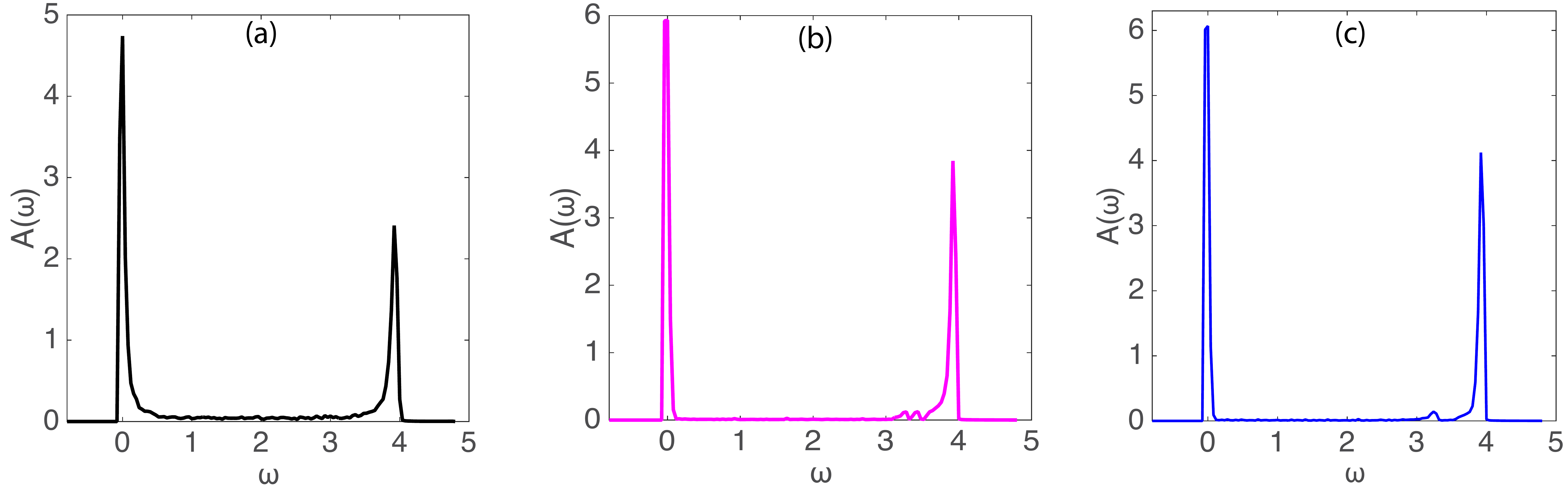}

\protect\caption{Disorder-averaged spectral functions. (a) The Anderson model without interaction. The parameters are chosen as $W=10J$ and $v_0=4J$, and the spectral function is averaged over $10^4$ random realizations. (b) The AA model without interaction. (c) The AA model with interaction ($U=J$). Other parameters are chosen as $\lambda=10J$ and $v_0=4J$, and $A(\omega)$ is averaged over $10^4$ different values of $\phi$ evenly distributed in $[-\pi,\pi]$. For all these three plots, the system size $L=20$.}\label{SupAwFreeInt}

\end{figure*}

\subsubsection{Calculations of the spectral function}

In the main text, we have defined the dynamic spectral function as 
\begin{eqnarray}
A(\omega)=\sum_{n'} |\langle G|n'\rangle|^2\delta(\omega-(E_{n'}-E_G)).\label{Spfun}
\end{eqnarray}
A direct calculation of $A(\omega)$ using Eq. \ref{Spfun} would also require a full diagonalizations of the many-body Hamiltonians to obtain the eigenstates $|n'\rangle$. We can avoid this by using the so called kernel polynomial method (KPM) \cite{WeiSe2006The} to calculate $A(\omega)$ for system size up to $L=24$. For each random realization of $H_I$, we  use the Lanczos algorithm to find the ground state $|G\rangle$ and the corresponding eigen-energy $E_G$. We then define a new Hamiltonian $H'=H_F-E_G$ to shift the eigen energies of $H_F$ and follow the standard procedures of KPM (see Ref. \cite{WeiSe2006The} for details). In order to fit the spectrum into the interval $[-1,1]$  (within which the Chebyshev polynomials are defined)
we apply a simple linear transformation to $H'$:\begin{eqnarray}
\tilde{H'}= (H'-b)/a,
\end{eqnarray}
where $a=(E'_{\text{max}}-E'_{\text{min}})/(2-\epsilon)$ and $b=(E'_{\text{max}}+E'_{\text{min}})/2$; Here  $\epsilon$ is a small cutoff parameter introduced to avoid stability problems ($\epsilon=10^{-3}$ is chosen in our numerical calculations); $E'_{\text{max}}$ and $E'_{\text{min}}$ are the maximal and minimal eigen energies of $H'$, respectively. Both $E'_{\text{max}}$ and $E'_{\text{min}}$ can be obtained by directly diagonalizing $H'$  using Lanczos algorithm. Then the spectral function can be expanded in Chebyshev polynomials \cite{WeiSe2006The}
\begin{eqnarray}
A(\tilde{\omega})=\frac{1}{\pi\sqrt{1-\tilde{\omega}^2}}\left(g_0\nu_0+2\sum_{k=1}^{K-1}g_k\nu_kT_k(\tilde{\omega})\right),
\end{eqnarray}
where $T_k(x)=\cos[k\arccos(x)]$ are the Chebyshev polynomials and $g_k=[(K-k+1)\cos\frac{k\pi}{K+1}+\sin\frac{k\pi}{K+1}\cot\frac{\pi}{K+1}]/(K+1)$ are the coefficients of the Jackson kernel; the moments $\nu_k=\langle G| T_k(\tilde{-H'})|G\rangle$ can be obtained iteratively based on the recursion relations of the Chebyshev polynomials 
\begin{eqnarray}
T_0(x)=1,\; T_{-1}(x)=T_1(x)=x,\; \text{and}\; T_{k+1}(x)=2xT_k(x)-T_{k-1}(x).
\end{eqnarray}
After $A(\tilde{\omega})$ is obtained, one can do a rescaling $\tilde{\omega}\rightarrow \omega=a\tilde{\omega}+b$ to obtain $A(\omega)$. In Fig. 4 (b) in the main text, we have plotted the disorder-averaged spectral function for the interacting Anderson model. In this plot, we have $L=20$ and used $10^4$ random realizations. The spectral function is normalized according to $\int A(\omega) d\omega=1$. As discussed in the main text, we have found unique features related to the exponential StOC in localized systems.  In Fig. \ref{SupAwFreeInt}, we also plot the spectral functions for the noninteracting Anderson and interacting AA models. Distinctive features of double peaks and exponentially decaying tails are also obtained in these cases.

\end{widetext}

\end{document}